\documentclass[conference]{IEEEtran}
\IEEEoverridecommandlockouts
\usepackage{amsmath,amsfonts}
\usepackage{algorithmic}
\usepackage{algorithm}
\usepackage{array}
\usepackage{float}
\usepackage{hyperref}
\usepackage[font=footnotesize 
]{subfig}
\usepackage{textcomp}
\usepackage{url}
\usepackage{verbatim}
\usepackage{graphicx}
\usepackage{caption}
\usepackage{subcaption}
\usepackage{cuted}
\usepackage{cite}
\usepackage{color, colortbl}
\definecolor{Gray}{gray}{0.9}
\usepackage{mathtools}  
\usepackage{amssymb}  
\usepackage{booktabs}
\usepackage[table, svgnames]{xcolor}
\usepackage{cellspace}
\usepackage{etoolbox}
\colorlet{headercolour}{DarkSeaGreen!40}
\usepackage{makecell}
\renewcommand{\theadfont}

\AtBeginEnvironment{tabular}{\rowcolors{1}{\ifnumless{\rownum}{2}{headercolour}{white}}{}}
\usepackage{lipsum}     

\begin{document}
\title{THEAS: Efficient Power Management in Multi-Core CPUs via Cache-Aware Resource Scheduling}

\author{\IEEEauthorblockN{1\textsuperscript{st} Said Muhammad}
\IEEEauthorblockA{\textit{Software and IT Engineering} \\
\textit{Ecole de Technologie Superieure}\\
Montreal, Canada\\
said.muhammad.1@ens.etsmtl.ca}
\and
\IEEEauthorblockN{2\textsuperscript{nd} Lahlou Laaziz}
\IEEEauthorblockA{\textit{Software and IT Engineering} \\
\textit{Ecole de Technologie Superieure}\\
Montreal, Canada\\
laaziz.lahlou.1@etsmtl.net}
\and
\IEEEauthorblockN{3\textsuperscript{rd} Nadjia Kara}
\IEEEauthorblockA{\textit{Software and IT Engineering} \\
\textit{Ecole de Technologie Superieure}\\
Montreal, Canada\\
nadjia.kara@etsmtl.ca}
\and
\IEEEauthorblockN{4\textsuperscript{th} Phat Tan Nguyen}
\IEEEauthorblockA{\textit{Architecture \& Technology R\&D} \\
\textit{Ericsson}\\
Montreal, Canada\\
Tan.Phat.Nguyen@ericsson.com}
\and
\IEEEauthorblockN{5\textsuperscript{th} Timothy Murphy}
\IEEEauthorblockA{\textit{Architecture \& Technology R\&D} \\
\textit{Ericsson}\\
Montreal, Canada\\
timothy.murphy@ericsson.com}
}


\maketitle
\begin{abstract}
Power optimization is a critical issue in modern CPU architectures, requiring innovative strategies to reduce energy consumption without compromising performance. Considering the high cost and time associated with hardware implementation, architectural simulators, such as the Generalized Event-driven Simulator with m5 (Gem5), are found to be the best alternative, providing an excellent platform for validating and evaluating power management schemes before deployment on real hardware. This research focuses on power optimization in multi-core CPUs by deploying a novel technique, Task Heterogeneity and Energy-Aware Scheduling (THEAS) algorithm, with an empirical power model, while setting per-core performance levels, targeting non-pinning (i.e., without assigning to specific cores) workloads.  In experimental analysis, the primary focus is on the Advanced Reduced Instruction Set Computing Module (ARM) architecture, specifically the Cortex-A72 model, by conducting full-system (FS) simulations using Gem5. Different algorithms of the Splash benchmark (CPU- and memory-intensive) are run as workloads on  Gem5 and real hardware, with similar configurations of the architecture. The research carried out provides information on power management and optimization techniques for ARM architecture-based systems, identifying areas for improvement in both hardware design and dynamic power management strategies. Promising results have been revealed by the proposed THEAS algorithm,  showing an error of approximately 5.6\% between Gem5 and Cortex-A72 real hardware-based simulation results. The error without THEAS is found to be approximately 10.4\%, resulting in an improvement of 4.8\% with THEAS. 
\end{abstract}
\begin{IEEEkeywords}
Gem5, THEAS, Non-pinning, ARM, Empirical power model.
\end{IEEEkeywords}

\section{INTRODUCTION}
\IEEEPARstart{M}{ulticore} systems are becoming essential in the digitalized environment, particularly in areas like scientific simulations, data processing, and parallel computing due to the increasing demand for high-performance and efficient computation in these domains.  The increasing demand for application processing from diverse users, including enterprises, researchers, and developers has driven a significant load on the cloud service providers.  Users require the execution of heterogeneous workloads, ranging from CPU-intensive and memory-intensive tasks to mixed workloads (combining CPU and memory-intensive tasks), which raises the need for high-performance computing devices (HPCs) with greater computational capabilities. However, this surge in the application processing demand and the development of HPC devices comes at the cost of increased power consumption, emerging as a critical concern of power management in modern multicore computing systems. \par
Consequently, efficient power management in multicore architecture is critical, especially for workloads requiring multiple cores stressing, leading to significant power consumption.  As a result, a large number of system-on-chip (SoC) power estimation tools such as Multicore Area Power Time (McPAT) \cite{Jianwang}, Computer-Aided Circuit Tool for Integrated-circuit (CACTI) \cite{Thasneema} and Wattch \cite{Brooks} have been developed by the researchers to assess the diverse power consumption estimation strategies which fall into one of the two categories, bottom-up, including McPAT, Cacti, and Wattch, and the top-down, which
utilizes the CPU parameters such as performance monitoring
counters (PMCs)  \cite{Hebbar, Mazzola}. These tools are specifically designed to simulate power and energy consumption based on CPUs, memory, and caches for different Infrastructure Set Architectures (ISAs), such as ARM, x86, and Reduced Instruction Set Computing (RISC-V).
\par
In Bottom-up Models (e.g., CACTI, McPAT, Wattch), the model utilizes the CPU's specifications (e.g., number of pipelines, cache size) to evaluate power utilization. It estimates the CPU complexity, area, component count, and the switching activity required for different tasks. McPAT is utilized as an integrated modeling tool to measure the power, area, and timing \cite{Zhai}. Whereas Cacti is a modeling tool for estimating memory and cache parameters, such as access time, cycle time, area, leakage power, and dynamic power.  The Wattch tool is a framework designed for power simulation at the architectural level. \par
 \IEEEpubidadjcol   
\IEEEpubidadjcol
The top-down strategy is based on CPU architectures and utilizes performance monitoring counters (PMCs) in conjunction with an empirical power model. PMCs are registers used to store the software and hardware data \cite{MJ}. While utilizing the PMCs, the empirical power model utilizes both architectural and microarchitectural events, such as instructions per cycle (IPC), L2 cache miss rate, and further speculatively executed instructions, including overall data access, overall misses, and overall hits. Based on this data, the empirical model provides a detailed analysis to help understand the behavior of the CPU and their role in the power consumption (both static and dynamic). Other research works use this data to train regression-based models to estimate the CPU power consumption \cite{Hebbar, Mazzola}.
\par
Collecting data from the hardware can be a challenging task due to several factors, such as the availability of the hardware platform that supports PMC monitoring,  which is often hard to collect. The available top-down approaches are less flexible than bottom-up approaches because they are designed specifically for a particular CPU or GPU architecture. When considering accuracy, the top-down approach outperforms the bottom-up approach (McPAT, Cacti, Wattch) since PMCs provide accurate data for various CPU architectures. 
\par 
To effectively implement the aforementioned strategies, selecting an architectural simulator proves to be a more viable choice, as it offers a cost-effective and efficient solution for evaluating various ISAs, providing detailed insights compared to real hardware. Such a simulator is commonly employed to assess the performance of a Central Processing Unit or system on chip (CPU/SoC), such as measuring the execution time, power, and low-level cache (LLC) performance,  during workload execution. Further, the ISAs-based tools such as Quick emulator (Qemu), SIMulation and Instruction-level Control System (Simics), Scalable Network-Infrastructure Performance Evaluation and Reporting (Sniper), and gem5 support the simulation of hardware-based events such as the number of L1, and L2 cache instructions, cache miss rates, and data cache access. The simulation of PMC events using ISAs helps to identify and understand performance bottlenecks in the system. ISAs facilitate design space exploration (DSE), allowing flexibility in configuring various hardware parameters such as the number of cores, cache sizes, and other architectural features.
\par
The architectural simulator GEM5 supports both bottom-up and top-down strategies to estimate the CPU's power consumption. A brief overview of the GEM5 simulator is given in section III. After selecting the simulator, the development of the scheduling technique is a crucial and challenging task due to traditional scheduling approaches, which often overlook workload heterogeneity (CPU-intensive, memory-intensive, and mixed workloads) and its impact on energy efficiency, leading to inefficient power management in multicore systems. To tackle the issue, PMCs can be helpful because they provide real-time insights into critical performance parameters such as IPC, cache miss rate, and branch mispredictions, which directly influence dynamic power consumption. 
\par
The key contributions of this research work are as follows, 

\begin{itemize}
\item To develop the  Task Heterogeneity Energy Aware Scheduling (THEAS) algorithm utilizing the PMCs events, such as IPC, cache miss rate. Integrate the THEAS algorithm into Gem5 and evaluate its performance using an empirical power model.
\item To analyze the impact of THEAS with core-based resource scheduling on CPU power consumption and validate the PMCs using an empirical power model on simulated and real ARM architecture.
\item To perform a comparative analysis of THEAS with standard GEM5 (without THEAS).

\end{itemize}

The research is further organized as follows: Section II discusses related work, while Section III defines the modeling and implementation framework for THEAS, the empirical power model, and Gem5. Section IV discusses the results and compares them with the real hardware for the validation, while  Section V  concludes the paper. 

\section{Related Work}

A light and accurate power modeling approach for heterogeneous computing platforms, specifically targeting CPU and GPU subsystems while utilizing PMCs for Dynamic Voltage and Frequency Scaling (DVFS) is developed by Sergio and Thomas et al. \cite{Sergio}. 
The author employs a systematic statistical approach to select minimal subsets of PMCs correlating with power dissipation, followed by training lightweight linear models for each subsystem across various frequencies. The proposed model demonstrates an average error of no more than 3.1\% when validated on real hardware, such as the NVIDIA Jetson AGX Xavier.  In addition, multicore systems face challenges in balancing performance scaling with energy efficiency, requiring optimal energy usage techniques, such as dynamic voltage and frequency scaling (DVFS) \cite{Mahmoudi}.  
\par

In \cite{Austin}, the author introduced SimpleScalar, a collection of instruments designed for CPU performance analysis as well as micro-architectural modeling supporting various ISAs, e.g., ARM, Alpha (RISC-based), Princeton Instruction Set Architecture (PISA), and x86. MARSSx86 \cite{Patel}, a qemu-based simulator, is developed with enhanced performance in terms of fast emulation of x86 architectures. The simulator uses the Dynamic Binary Translation (DBT) technique, executing the tasks with minimal time-delay tradeoffs.  In \cite{Binkert}, the author introduced Gem5, a versatile and adaptable platform supporting various ISAs, including ARM, RISC-V, x86, Power, Null, and Scalable Processor Architecture (SPARC).  Yahya and Magnus et al. developed a fast and accurate energy model with DVFS implementation in the Gem5 simulator, specifically tailored for edge computing applications with stringent power constraints \cite{Yassin}. The author proposed a non-intrusive, application-controlled DVFS management system, with the limitation of targeting only System Call emulation (SE) mode in Gem5.
\par
Butko et al. \cite{Butko} considered the Gem5 and McPAT to emulate the ODROID-XU3 board for the experimental analysis, which is a heterogeneous architecture consisting of ARM Cortex-A72 and Cortex-A53.  The author reported an average error of 24\% between the measured energy from hardware and the energy measured by Gem5 with McPAT. Using gem5, a top-down strategy has been deployed by Basireddy et al. \cite{Reddy}, which implements the empirical power model on the ARM Cortex-A15. The author obtained an error of less than 6\% with 60 workloads, targeting the ARM heterogeneous architecture with big-little cores. Furthermore, with the reduction of workloads from 60 to 15, the error increased approximately by 100\%, reaching an overall error of 10\% from 6\%, which provides an opportunity for further research with fewer workloads.
\par
Y. Qiu et al. focused on evaluating the accuracy of the Gem5 simulator in modeling modern ARM server architectures, highlighting the importance of simulator precision for credible research outcomes \cite{Qiu}. The authors employ a systematic methodology to measure performance errors quantified by the mean absolute percentage error (MAPE) metric. With single-core mode, the MAPE value is found to be 26.31\%, while in multi-core mode, it remains around 30\%  for specific benchmarks, providing insights into the performance discrepancies between the simulator and the actual ARM processor. 
\par
In light of the details above, Gem5 is found to be the most suitable simulator for architectural simulation due to its enhanced flexibility in terms of resource management and adaptability, which enables an efficient experimentation environment for prototyping and optimization across diverse system configurations. 
\par   
In this research,  the empirical power model is inspired by the novel power model developed by Walker et al. \cite{Bischoff}, which was designed for the heterogeneous big.LITTLE architecture of ARM ISAs. We adapt and apply it to a single-ISA architecture, modifying it by using PMC events such as IPC, cache miss rate, overall access, and simulation time. The author in \cite{Bischoff} obtained various performance levels, where the mean absolute percentage error (MAPE) between the real hardware and simulation environment is found to be 6.6\% for the Cortex-A7, while 3.3\% for Cortex-A15 using the chip selection instead of cores. The experiment was conducted with 65 workloads, whereas implementing the power model on a single, customized architecture with a smaller number of workloads remains a challenging task.  \par

Considering the challenges arising in \cite{Reddy} and \cite{Bischoff}, we targeted fewer workloads on a single ARM Cortex-A72 chip. The aim is to first develop an adaptive technique that can choose a core based on the workload computation demand and integrate it with the empirical power model in Gem5 to analyze the scalability and precision of the model.

\section{Proposed Approach Implementation Framework}
To develop the THEAS algorithm and empirical power model, the PMC events, such as IPC, overall cache miss rates, and performance levels based on simulation time, frequency, and voltage, are considered for integration into Gem5 by adapting the parameters of the ARM Cortex-A72.  The implementation framework consists of the simulator environment setup, performance adaption mode in Gem5, the proposed algorithm, and the empirical power model.
\begin{figure}[h!]
\centering
\includegraphics[width=3.55in, height=2.6in]{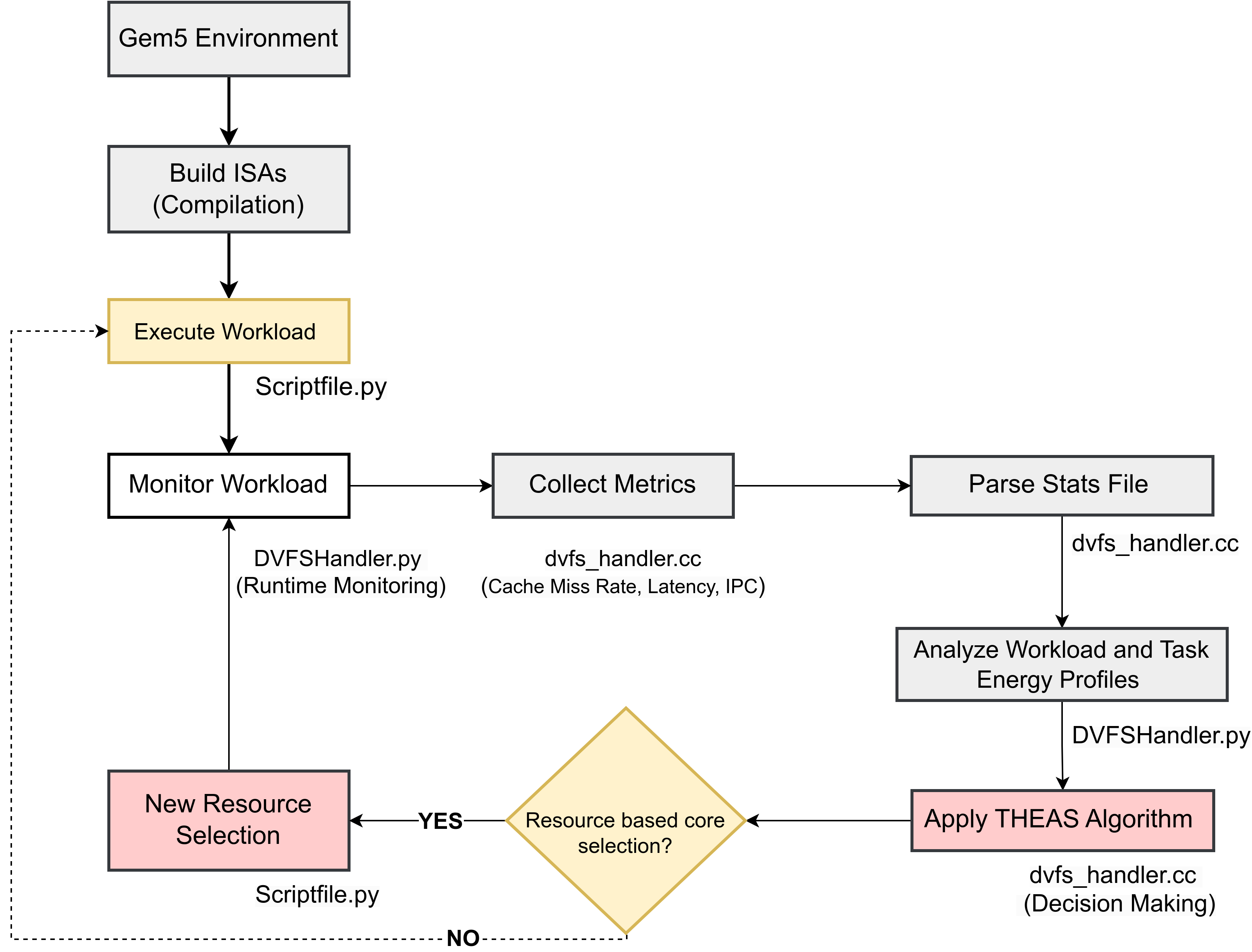}
	\caption{Retrieval and deployment of PMC events along with THEAS and core-based resource level on Gem5.}
\end{figure}
The modeling strategy employed in this research work is illustrated in Fig. 1, which utilizes the Gem5 23.0 version with the ARM architecture.  Considering the runtime-based scheduling, DVFS handler, chip selection, and caches are found to be the most critical parameters to modify to tackle the proposed objective of handling power consumption issues in multicore CPUs with unassigned workloads to CPU cores (Non-pining).  A brief description of the Gem5 setup environment, followed by core-based resource allocation and the THEAS algorithm, is given in the following subsections.  \par  
\subsection{Gem5 setup and its Evaluation}
This section provides a brief overview of the Gem5 simulator \cite{Binkert}, followed by its usage as an experimental environment in the proposed research. 

\subsubsection{Gem5 CPU model }
Gem5 supports several CPU models, including minor, Timing, Atomic, out-of-order (O3), high-performance in-order (HPI), Null, and power \cite{NL}. 
\par
In this research work,  Out-of-order (o3) CPU is selected because of its wide support of configuration, full system simulation, and enhanced resource utilization capability. The general architectural parameters shown in Table II serve as the basis for configuring the simulation environment operating in full system mode with the ARM Cortex-A72.  The PMCs' events, such as instruction per cycle, data cache miss rate, and simulation time, are used in conjunction with the frequency and voltage. Furthermore, the L1 and L2 caches are modeled using the PMCs' parameters, which reflect the memory hierarchy of the Cortex-A72.
\subsubsection{Accuracy evaluation }
The accuracy evaluation process involves experimenting on both the Gem5 simulator and real hardware with matching configuration and workload (Splash benchmark) as depicted in Fig. 2. An empirical modeling scheme is used in the Gem5 to collect power consumption by using the PMCs-based event and compare it with the real hardware. 
\begin{figure}[h!]
	\centering
	\includegraphics[width=1.9in, height=2.2in]{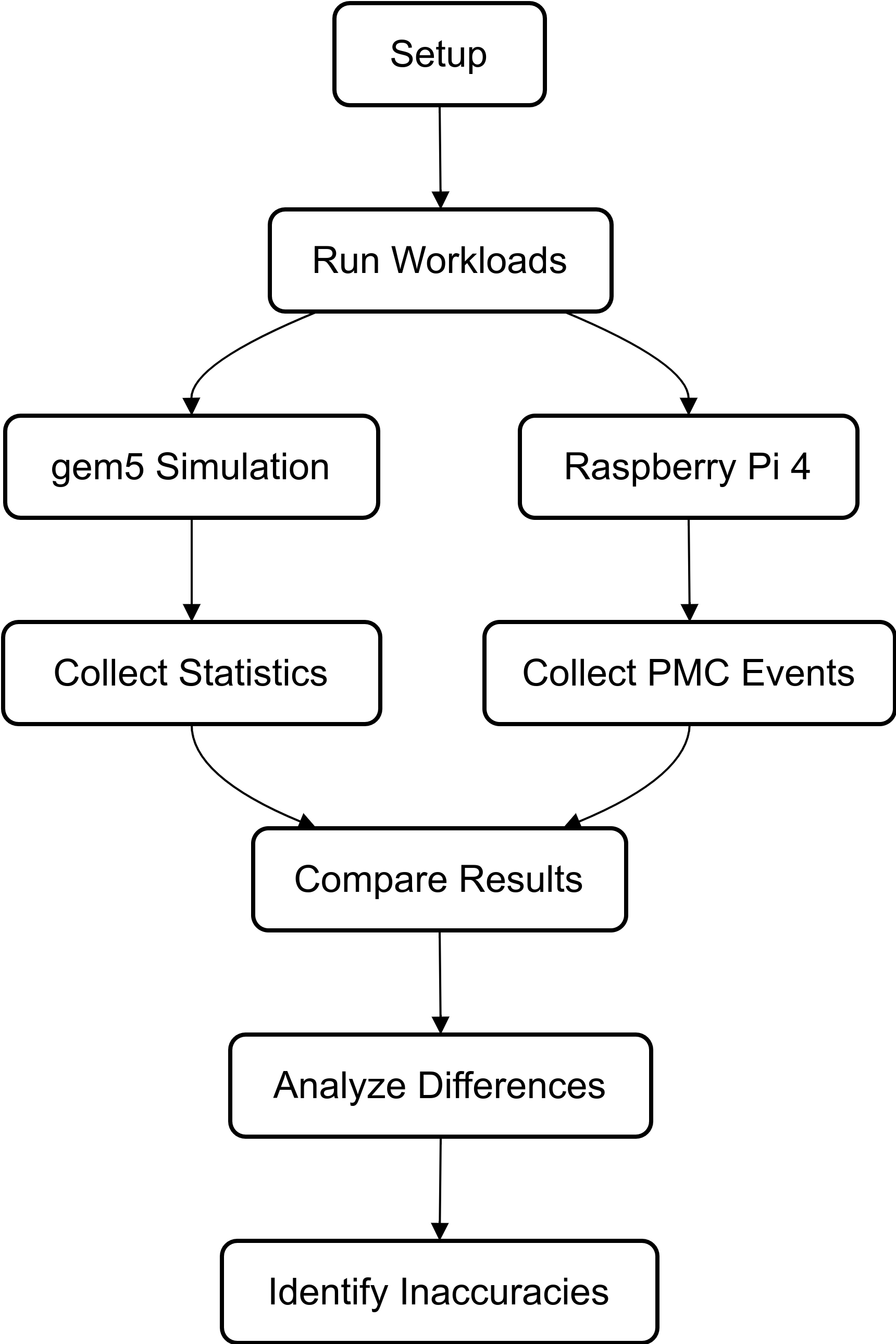} 
	\caption{Gem5 with real  hardware accuracy evaluation process.}
\end{figure}
The most prominent PMCs events, used in the modeling scheme to collect the power consumption during the simulation, are shown in Table 1.   
\subsection{Core-based resource adaption in Gem5}
The core-based performance adaptation mechanism in the Gem5 is outlined in Fig. 3, to schedule threads based on non-pinning tasks dynamically. The core selection in Gem5 is initiated by analyzing workload statistics and workload energy profiles to evaluate the current power and performance metrics.
\begin{figure}[!htp]
	\centering
	\includegraphics[width=3.0in, height=1.3in]{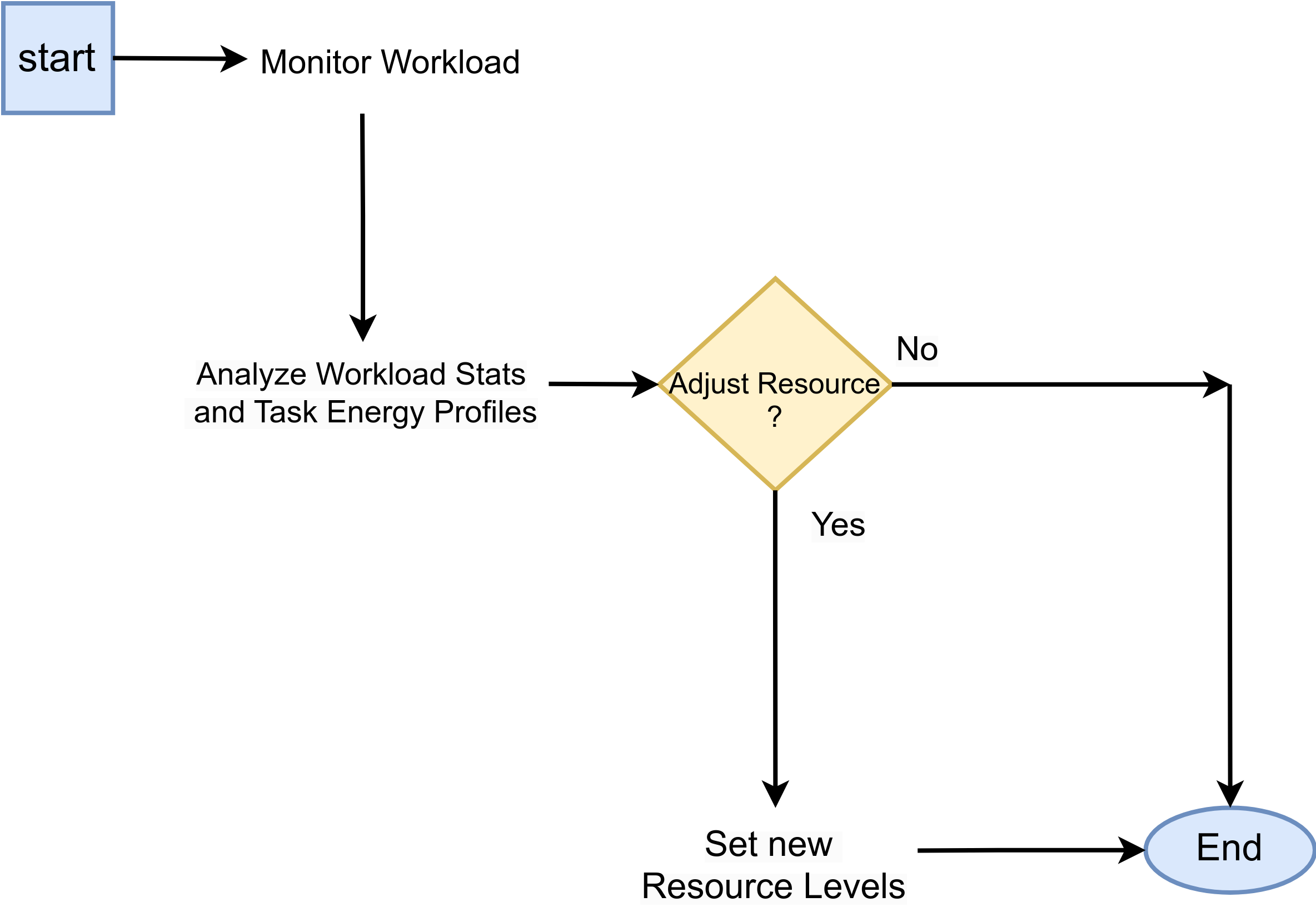}
	\caption{Cores adaptation for dynamically scheduled threads in a non-pinning-based workload.}
\end{figure}
A decision block checks the system behavior to determine if any resource level adjustment (LOW, MEDIUM, or HIGH threshold) is required based on the workload characteristics. If no adjustment is necessary, the system maintains the current resource level, however, during the need for resource adjustment, new resource levels are set, reflecting changes in frequency and voltages to optimize the power consumption. 
\begin{table}[!htp]
\centering
\caption{PMC events used in Empirical Power Model}
{\fontsize{7pt}{7pt}\selectfont
\begin{tabular}{|l|l|l|}
\hline
\textbf{Abbreviation} & \textbf{Parameter} & \textbf{Unit} \\ \hline
voltage & Voltage & V (Volts) \\ \hline
IPC & Instructions per Cycle & Unitless \\ \hline
dcache\_overallMisses & Data Cache Overall Misses & Misses \\ \hline
simSeconds & Simulation Time & Seconds \\ \hline
overallAccesses & Overall Accesses & Accesses \\ \hline
$P_{dyn}$ & Dynamic Power (Core) & W (Watts) \\ \hline
$P_{dyn,i}$ & Dynamic Power (L2 Cache) & W (Watts) \\ \hline
\end{tabular}
}
\end{table}
\subsection{THEAS algorithm}
Task Heterogeneity and Energy Aware Scheduling (THEAS) is a novel approach proposed in this research work alongside the core-based performance selection algorithm to tackle the non-pinning based workloads. The algorithm presented first time in this research work, addresses the dual challenges of task heterogeneity and energy efficiency in modern computing systems.

\begin{table}[!htp]
\centering
\caption{Cortex-A72 parameter set in Gem5}
{\fontsize{6.5pt}{6.5pt}\selectfont
\begin{tabular}{|l|l|}
\hline
\textbf{Parameter} & \textbf{Specification} \\ \hline
Core type & ARM Cortex-A72 (in-order, MinorCPU) \\ \hline
Cores & 4 \\ \hline
CPU clock (MHz) & 200, 800, 1200, 1800 \\ \hline
DRAM Size (MB) & 8192 \\ \hline
DRAM Clock (MHz) & 1600 \\ \hline
L2 Cache & Size: 1 MB, $~~$ Assoc.: 16, $~~$ MSHRs: 10  \\
         & Latency: 20-30 cycles\\
         & Write buffers: 16 \\ \hline
L1-I Cache & Size: 48 kB,  $~~$ Assoc.: 3, $~~$ MSHRs: 4 \\
           & Latency: 2 cycles\\ \hline
L1-D Cache & Size: 32 kB,  $~~$ Assoc.: 4, 
 $~~$MSHRs: 6  \\
           & Latency: 2 cycles, $~~$ Write buffers: 12 
           \\ \hline
ITLB/DTLB & 128 entries each \\ \hline
Branch predictor & TAGE \\ \hline
BTB entries & 4096 \\ \hline
RAS entries & 48 \\ \hline
ROB entries & N/A (In-order CPU) \\ \hline
IQ entries & N/A (In-order CPU) \\ \hline
Front-end width & 2 \\ \hline
Back-end width & 2 \\ \hline
LSQ entries & 16 \\ \hline
\end{tabular}
}
\end{table}
THEAS operates on the principle of continuously monitoring key performance metrics that serve as indicators of workload heterogeneity and system performance.  The core functionality of the THEAS algorithm, as described in Algorithm 1, consists of the following steps. 

\textit{Performance Metric Monitoring: }The algorithm continuously collects and analyzes PMCs, such as  IPC, cache miss rate, and fetch rate for each core. \\
\textit{Decision Making: }THEAS compares the collected PMCs against predefined thresholds, e.g., IPC and cache miss rate, to determine the appropriate course of action. \\
\textit{Resource Level Adjustment: }While utilizing the thresholds, the algorithm decides whether to change to a higher, lower, or maintain the current resource level-based core.\\
\textit{Core adjustment Implementation: }During the change in the resource level, such as the cache miss rate exceeding means, where the workload will be scheduled to a core (medium or low level) with less computation resource, while in the opposite case, the higher resource level core will be selected. While considering the IPC, a higher IPC means more instructions per cycle, demanding the selection of a core with higher computation capability, while a low IPC indicates lower computation demand. 
\begin{algorithm}[htp]
\small
\caption{THEAS (Task Heterogeneity and Energy Aware Scheduling) Algorithm}
\begin{algorithmic}[1]
    \STATE \textbf{Input:} Core performance metrics (IPC, cache miss rate, fetch rate), Current resource level
    \STATE \textbf{Output:} Updated resource level and Frequency settings
    \STATE \textbf{Initialize:} $IPC\_threshold\_low, IPC\_threshold\_high,$  \\  $cache\_miss\_threshold, fetch\_rate\_threshold$
    \WHILE{system is running}
        \FOR{each core}
            \STATE Read performance statistics from the file
            \STATE Extract IPC, cache miss rate, and fetch rate
            \IF{$IPC < IPC\_threshold\_low$ \AND $cache\_miss\_rate > cache\_miss\_threshold$}
                \STATE $new\_level \gets \min(current\_level + 1, HIGH)$
            \ELSIF{$IPC > IPC\_threshold\_high$ \AND $fetch\_rate > fetch\_rate\_threshold$}
                \STATE $new\_level \gets \max(current\_level - 1, LOW)$
            \ELSE
                \STATE $new\_level \gets current\_level$
            \ENDIF
            \IF{$new\_level \neq current\_level$}
                \STATE Adjust frequency and voltage for the core
                \STATE Update current resource level
            \ENDIF
        \ENDFOR
        \STATE Wait for the next scheduling cycle
    \ENDWHILE
    \STATE \textbf{End}
\end{algorithmic}
\end{algorithm}
\noindent
The dynamic adaptation of resource levels enables the system to enhance energy efficiency while maintaining the necessary computational resources, particularly in scenarios where workloads fluctuate significantly over time. The proposed approach can play a crucial role in heterogeneous systems where workload characteristics are not uniformly distributed, such as non-pinning tasks. The deployed THEAS algorithm in this research work ensures a balance between performance and power consumption, making it suitable for a wide range of real-time applications. A comparative analysis of the proposed THEAS algorithm with well-known scheduling techniques such as Completely Fair Scheduler (CFS), Energy-Aware Scheduling (EAS), Heterogeneous Scheduling (HeteroSched), and Utility-Based Scheduling is presented in Table III. Each scheme is compared based on adaptability, core selection criteria, performance scaling, cache awareness, overhead, and real-time suitability.
\begin{table}[htbp]
\centering
{\small\captionsetup{font=small}\caption{THEAS comparison with available Scheduling Schemes}}
\includegraphics[width=\linewidth]{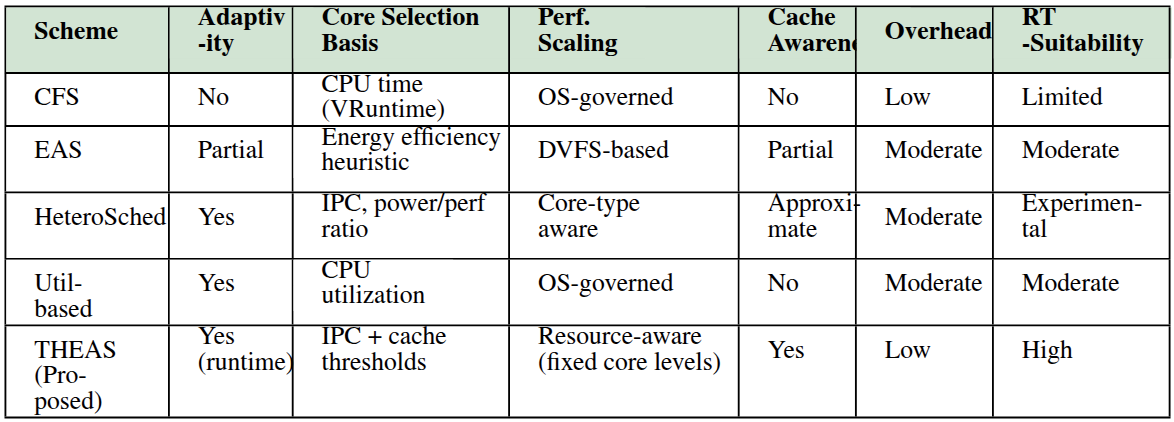}  
\label{tab:schemes}
\end{table}

\section{Experimental Setup, Results and Discussion}
This section provides details about the experimental validation of THEAS using an empirical power model with Gem5, as well as real hardware (Raspberry Pi 4), to assess the model's accuracy.

\subsection{ Empirical Power Model Evaluation Using Gem5}
Considering the accuracy level and scalability, the empirical power model is used to obtain power consumption details during runtime.  The coefficients of the empirical power model are briefly defined in Table I.  The empirical power model is integrated into the Gem5 simulator at both the CPU and L2 cache levels. \\
\textbf{CPU power model: }
The basic CPU power consumption model for dynamic power is derived from \cite{Rabaey},

\begin{equation}
    P_{dyn}^{cpu} = \alpha \times C \times V^2 \times f
\end{equation}
Where $\alpha$ represents the activity factor (fraction of transistors switching at any given time), C is the total capacitance being switched per clock cycle, V is the supply voltage, and $f$ is the clock frequency. Equation 1 is developed based on the physics of transistors, defined as the power consumed when capacitive loads are charged and discharged, depending on the voltage (squared), capacitance, and the rate at which it is integrated (frequency). 
While executing the empirical formula, the activity factor is transformed into the processor's Instruction Per Cycle (IPC), which measures the number of instructions executed per clock cycle. 

\begin{equation}
 P_\text{dyn}^{cpu} = V \times (2 \times \text{IPC}) 
\end{equation}
Equation 2 defines that the execution of IPC has a direct impact on power consumption with an adjustment constant, as used in the two, which is based on the observed behavior of the processor (showing that each instruction plays a significant role in the power consumption). \par
The simulation data obtained from the Gem5 stats.txt file indicates power consumption across different cores, implying that the workload is effectively used by multiple cores based on their computation demands, as expected by the benchmark.
\begin{figure*}[h!]
  \centering
  \begin{minipage}{0.44\textwidth}
    \centering
    \includegraphics[width=1.1\linewidth, height=0.28\textheight, keepaspectratio]{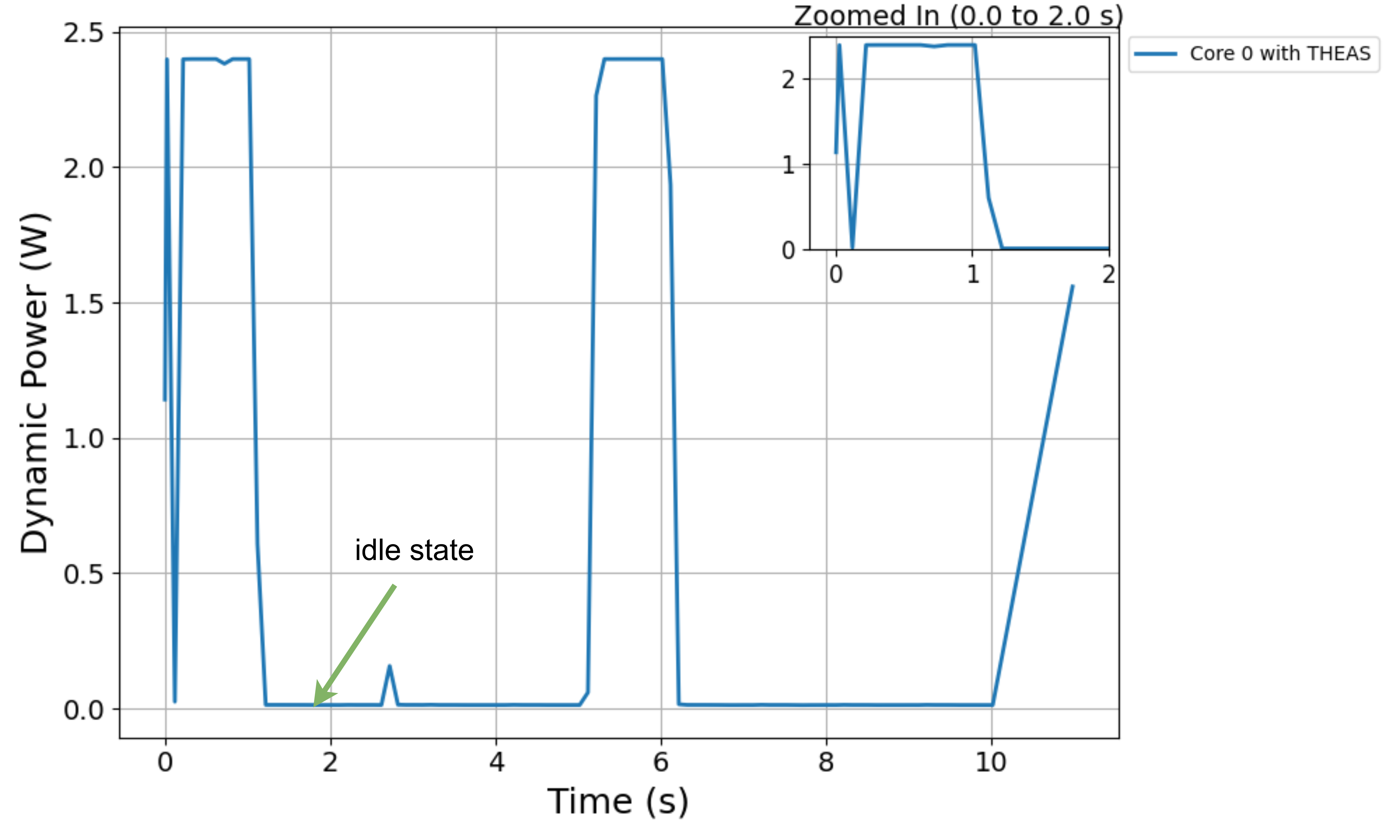}
    \caption*{\footnotesize(a) CPU 0 Dynamic Power}
  \end{minipage}%
  \hspace{0.03\textwidth}
  \begin{minipage}{0.44\textwidth}
    \centering
    \includegraphics[width=1.1\linewidth, height=0.28\textheight, keepaspectratio]{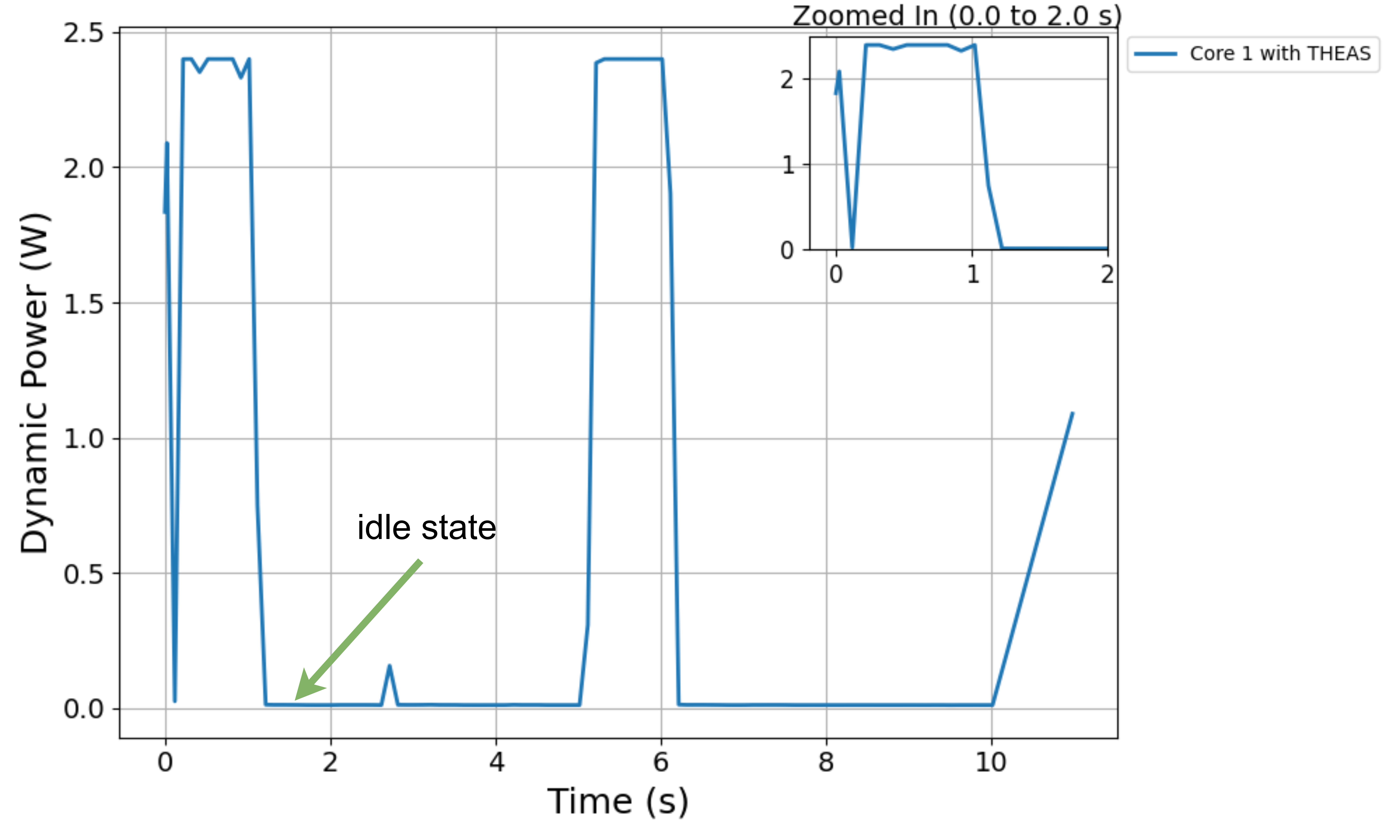}
    \caption*{\footnotesize(b) CPU 1 Dynamic Power}
  \end{minipage}
  \caption{Splash Barness benchmark per core power consumption vs time(s) steps.}
\end{figure*}
\begin{figure*}[h!]
  \centering
  \begin{minipage}{0.43\textwidth}
    \centering
    \includegraphics[width=1.11\linewidth, height=1.0\textheight, keepaspectratio]{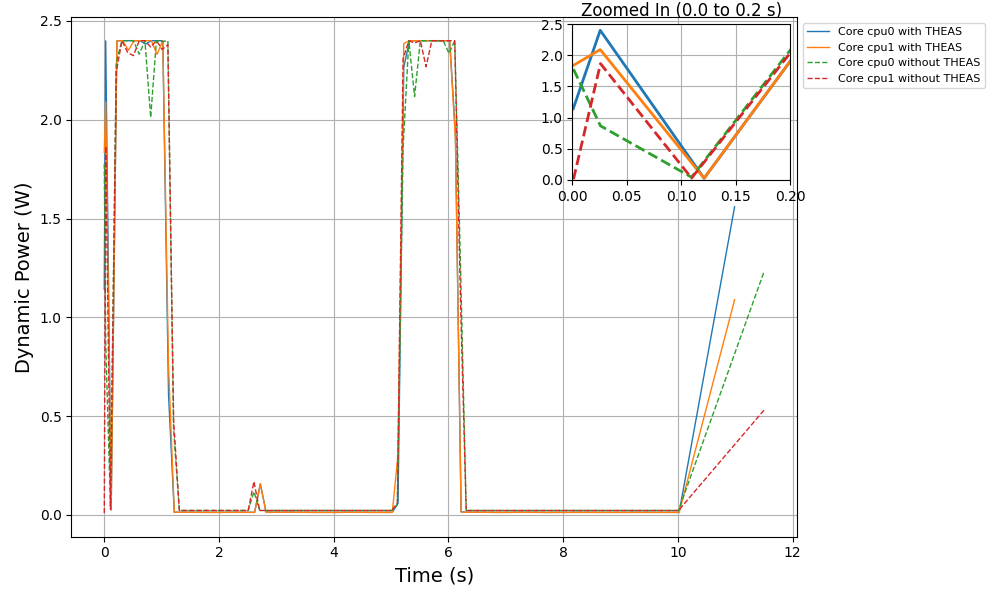}
    \caption*{\footnotesize(a) FMM execution on 2 cores with 8 processes}
  \end{minipage}%
  \hspace{0.04\textwidth}
  \begin{minipage}{0.43\textwidth}
    \centering
    \includegraphics[width=1.11\linewidth, height=1.0\textheight, keepaspectratio]{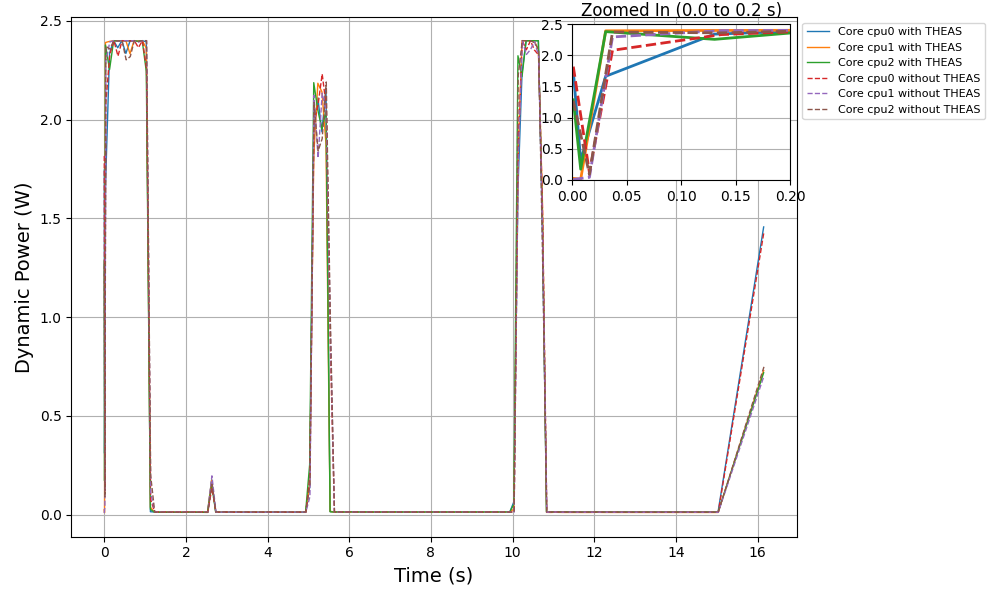}
    \caption*{\footnotesize(b) Mixed workload execution on 3 cores with 8 processes}
  \end{minipage}
  \caption{Dynamic power consumption of Splash benchmark algorithms for multicores.}
\end{figure*}
The final CPU and L2 cache power models used in gem5 \cite{Arm_powermodel} are shown in Equations 3 and 4, respectively.  
\begin{align}
P_{dyn}^{cpu} &= voltage \times \bigg(2 \times IPC \\  \nonumber
&+ 3 \times 10^{-9} \times \frac{dcache\_overallMisses}{simSeconds} \bigg)
\end{align}
\textbf{L2 cache power model: }
\begin{equation}
    P_{dyn,i}^{L2} = 0.000018\times\sum_{i=1}^{N}{overallAccesses}_{i}
\end{equation}
Equation 3 utilizes the IPC with overall cache miss rates and simulation time, whereas Equation 4 employs the overall accesses generated during the workload execution.
\par
In Gem5, experiments are conducted under resource levels HIGH, MEDIUM, and LOW, set at 1800MHz, 1200MHz, and 800MHz, respectively, with Cortex-A72 configurations.  In-depth details about the Cortex-A72 for the Gem5 environment are given in Table II. The CPU's dynamic power consumption while using the Splash benchmark Barnes algorithm as a workload is shown in Fig. 4. The graphs display the dynamic power consumption of two CPU cores (0 to 1) over time with eight processes in Gem5 using FS mode. Initially, to better elaborate on the THEAS impact, the workload is executed at two resource levels, HIGH and LOW. 
Both cores are attenuated at different levels of power consumption demand, adapting to the resource level set by the THEAS algorithm.  Core 0 is set with a low-frequency level (800 MHz), based on the low rate of the PMC policy parameters, which shows a start at a slightly lower power consumption rate (1.2 Watts). Core 1, on the other hand, started at 1.5 Watts, indicating the adaptation of the core with high computational capability. The idle state of the power consumption is achieved when the computation demand is not in the threshold regions or during the delay between consecutive workloads.  The transition obtained demonstrates that THEAS dynamically reduces resource levels as workload intensity decreases by selecting cores with lower frequency and voltage levels, optimizing energy usage.
\par
Fig. 5 shows the average dynamic power consumption over time for memory-intensive Fast Multipole Method (FMM) and mixed workload (FMM with Barnes), running each with eight processes simultaneously. Whereas Fig. 5(a) compares the dynamic power consumption for the FMM workload with and without THEAS using two cores (0, 1), a significant difference is observed in core invocation, with 1.2 watts and 2 watts. Without THEAS, core 1 starts at 0, showing that the cores are active while there is no workload. 
Furthermore, in Fig. 5(b), the mixed workload is plotted with three cores (LOW, MEDIUM, and HIGH), illustrating the adaptation of THEAS through core invocation. Following Fig. 5(a), in Fig. 5(b), the cores are also activated at different levels.
\begin{figure*}[htbp!]
  \centering
  \begin{minipage}{0.40\textwidth}
    \centering
    \includegraphics[width=1.0\linewidth, height=1.4\textheight, keepaspectratio]{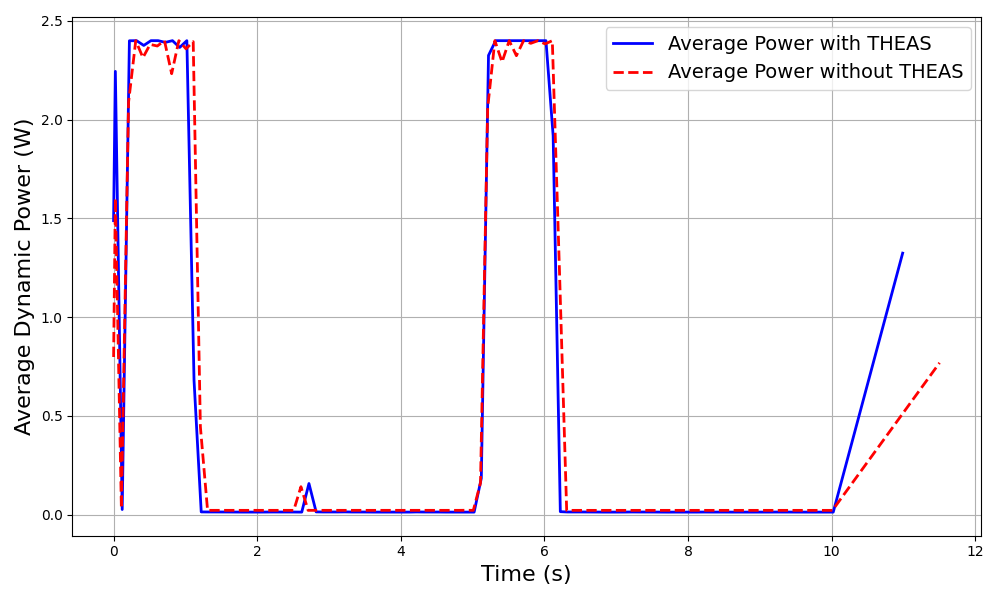}
    \caption*{\footnotesize(a) FMM execution on 2 cores with 8 processes}
  \end{minipage}%
  \hspace{0.05\textwidth}
  \begin{minipage}{0.40\textwidth}
    \centering
    \includegraphics[width=1.0\linewidth, height=1.4\textheight, keepaspectratio]{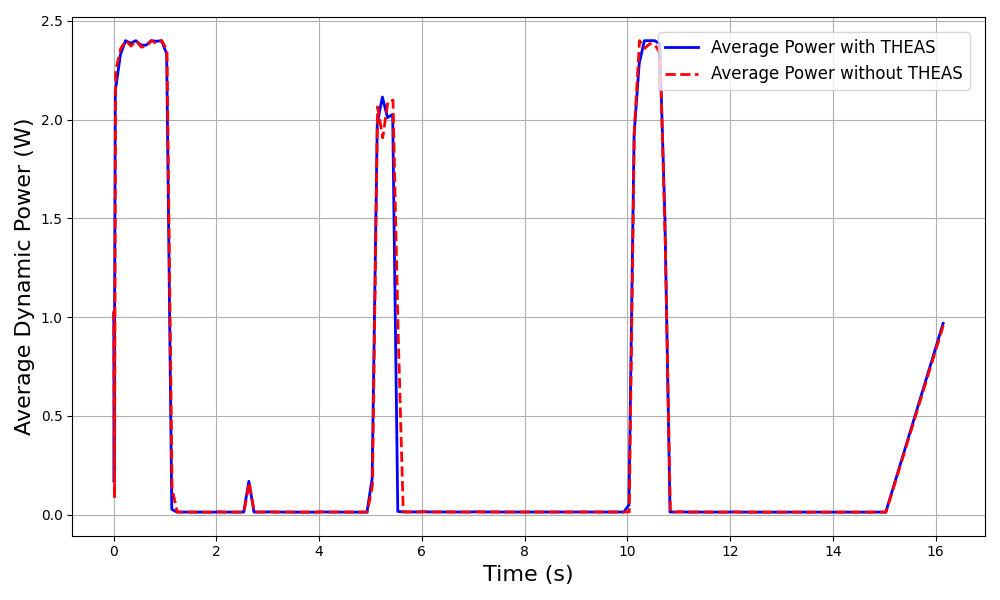}
    \caption*{\footnotesize(b) Mixed workload execution on 3 cores with 8 processes}
  \end{minipage}
  \caption{Average dynamic power consumption of Splash benchmark algorithms.}
\end{figure*}
\begin{figure*}[htbp!]
  \centering
  \begin{minipage}{0.42\textwidth}
    \centering
    \includegraphics[width=1.12\linewidth, height=1.4\textheight, keepaspectratio]{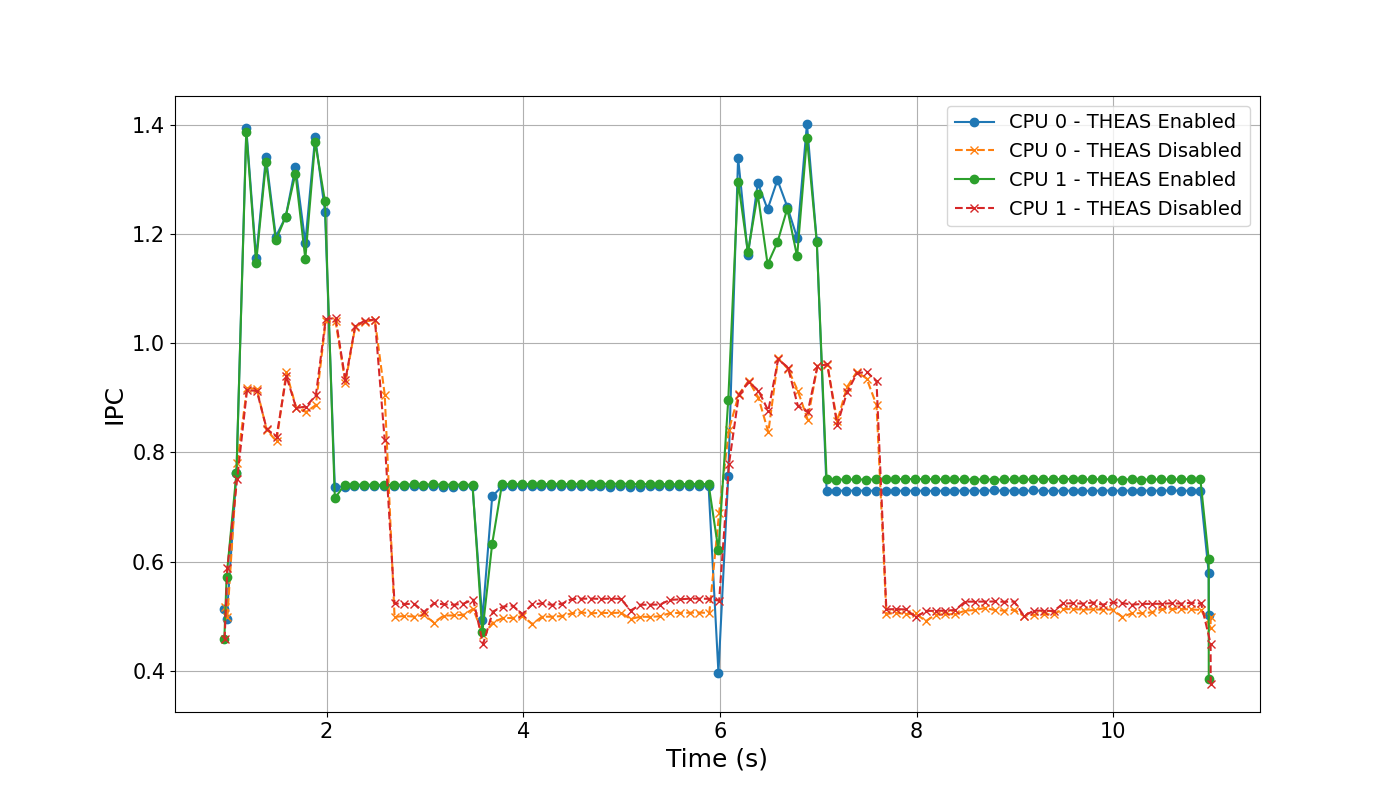}
    \caption*{\footnotesize(a) Splash  FMM execution on 2 cores with 8 processes}
  \end{minipage}%
  \hspace{0.03\textwidth}
  \begin{minipage}{0.42\textwidth}
    \centering
    \includegraphics[width=1.12\linewidth, height=1.4\textheight, keepaspectratio]{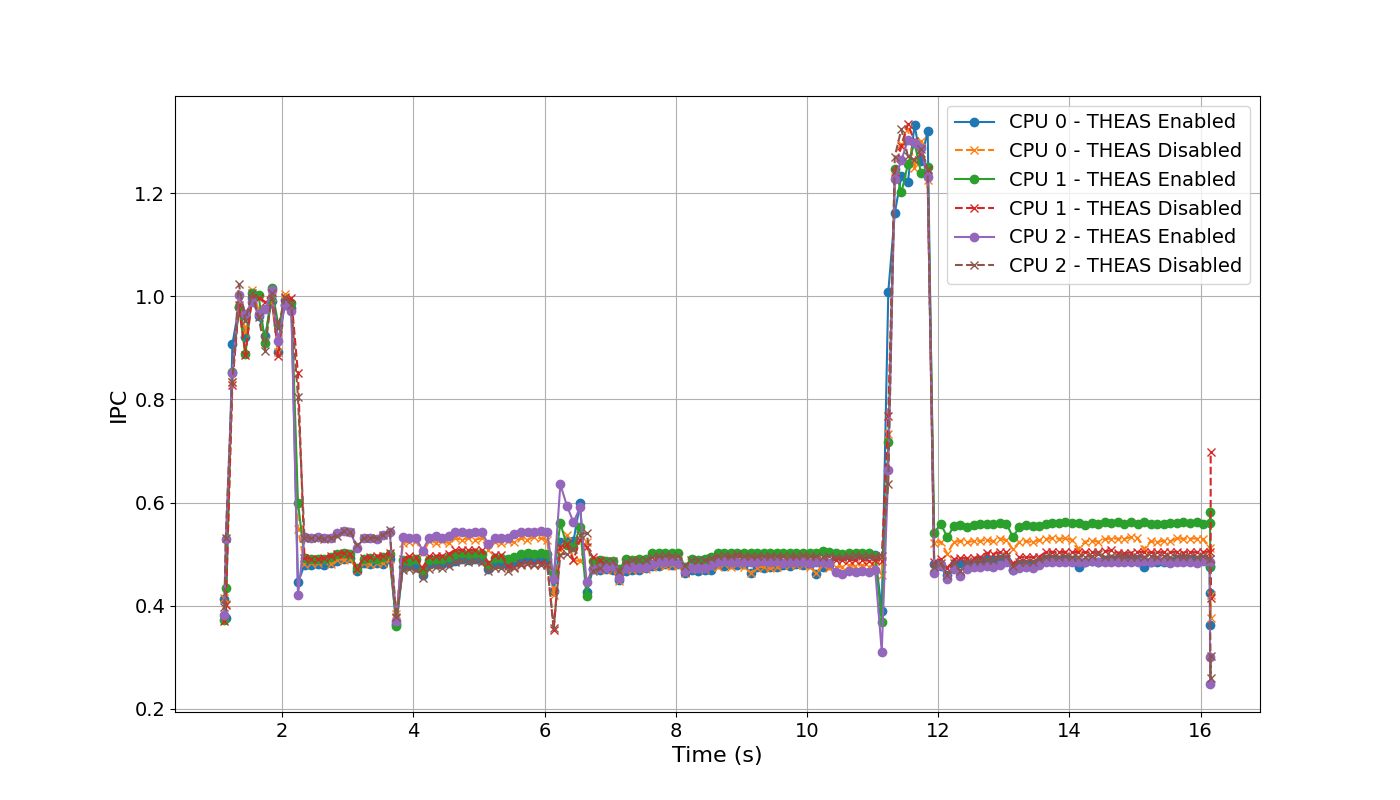}
    \caption*{\footnotesize(b) Mixed workload execution on 3 cores with 8 processes}
  \end{minipage}
  \caption{Splash benchmark IPC.}
\end{figure*}
In Fig. 6, the average power consumption is plotted for the FMM and mixed workloads to illustrate the difference between the THEAS and non-THEAS algorithms, highlighting the impact of THEAS on power consumption.
Considering FMM, an improvement of \textbf{4.64\% }by reducing dynamic power consumption from \textbf{0.530125 W} to \textbf{0.505525 W} without any performance degradation. While the mixed workload (6 (b)) dynamic power consumption is improved by \textbf{3.12\%}, reducing from \textbf{0.35356 W} to \textbf{0.34254 W}  without compromising on execution time. To observe the prominent difference with real hardware, FMM is further jointly executed with a mixed workload by doubling the number of processes. The average power consumption with THEAS reached up to \textbf{1.71613 W}, while without THEAS it was \textbf{1.80737 W}, contributing almost \textbf{5\%} to the reduction in power consumption.  
When workloads fall below the threshold or during delays between distinct workloads, IPC drops to zero, reflecting THEAS's dynamic management. The observed discrepancy between IPC and power consumption, where power initiation occurs at zero while IPC exhibits a slight temporal shift (approximately 1 second or less), stems from the inherent system initialization within gem5, which begins power modeling at zero. Additionally, the THEAS algorithm relies on periodic analysis of the stats.txt file, an output of gem5 execution, which introduces a processing latency that delays IPC-driven scheduling decisions.  Without THEAS, cores do not execute efficiently due to the absence of thresholds, resulting in longer execution times and increased energy consumption, as the cores remain consistently active even when the workload demands are lower. 
While in 7(b), CPU 0 and CPU 2 exhibit higher IPC bursts at different times, indicating that THEAS is dynamically shifting the workload to these cores based on the changing demands of the workload. In contrast, when THEAS is disabled, all three CPUs maintain a relatively lower IPC profile, suggesting a lack of dynamic resource allocation. The dynamic adjustments in IPC demonstrate the system's ability to adapt to varying computational demands, showcasing the benefits of energy-aware scheduling in heterogeneous workloads. The implied empirical model ensures these transitions are well-calculated, adjusting dynamically to the workload's requirements. 
\vspace{0cm}
\subsection{Workload validation on Real Hardware}
The results retrieved via THEAS and the empirical power model in the Gem5 are validated using real hardware, specifically choosing the Raspberry Pi 4 Model B  with 8GB LPDDR4 RAM due to its architectural similarity (ARM Cortex-A72). 
The real hardware validation is performed using a similar workload (the Splash benchmark) as in Gem5. The UMC25 power meter is used to measure voltage and current consumption as shown in Fig. 8. The power consumption from the real hardware is obtained by the power formula given in Equation 5 using real hardware statistics, e.g., voltage and current.
 \begin{figure}[!h]
	\centering
	 \includegraphics[width=2.6in, height=1.6in]{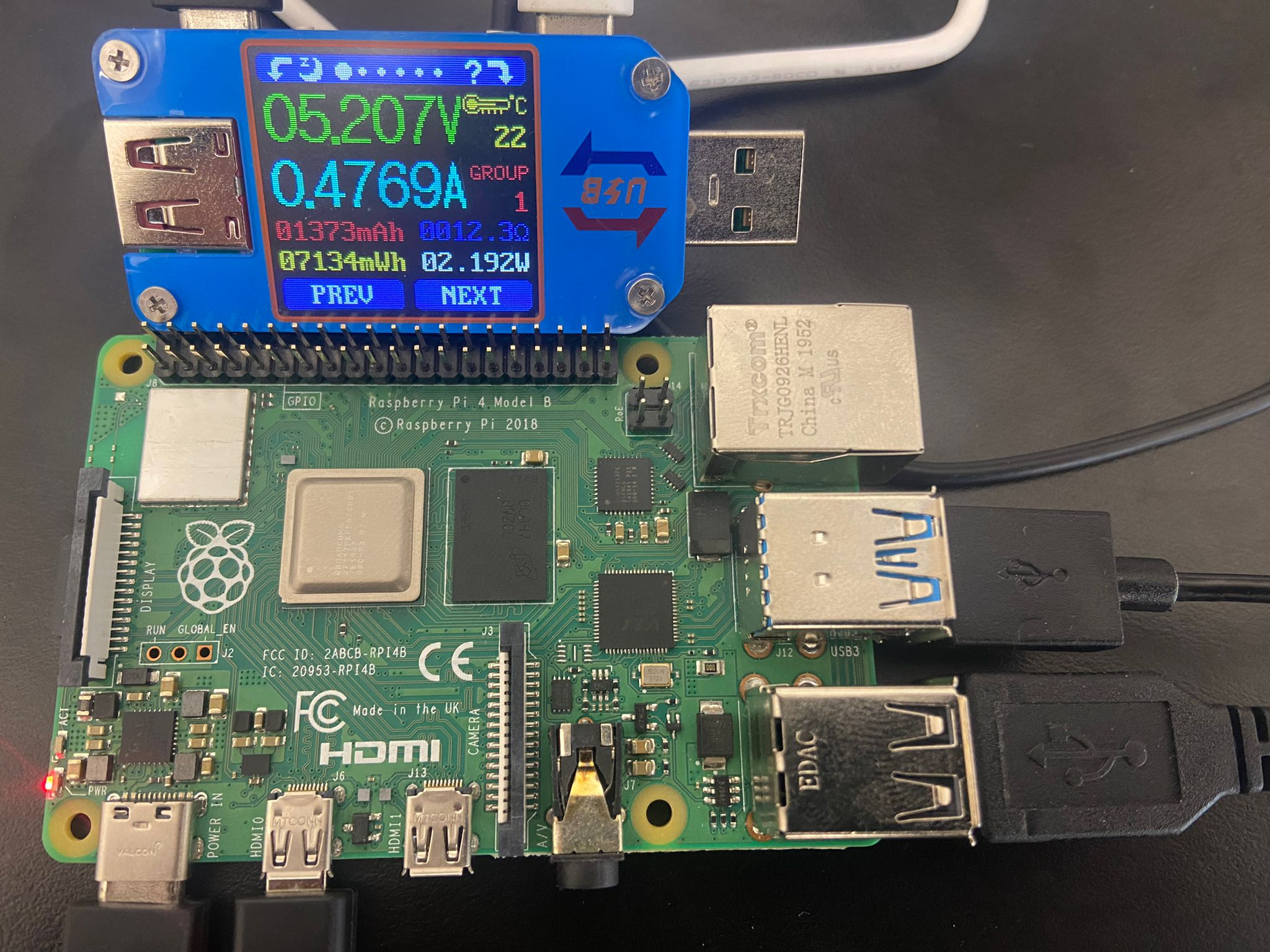}
  \caption{Hardware of ARM cortex-A72 (Raspberry Pi4).}
\end{figure}
In Equation 5, P$_{dyn}$ represents the dynamic power, V is the voltage, and I$_{dyn}$  is the current in amperes.  During the analysis of the experiment, the measured voltage on hardware was \textbf{5.207V} with minor fluctuations. While considering current, the measured values on real hardware provide a significant variation in current consumption, therefore multiple experiments have been conducted as shown in Table IV to obtain the average.
\begin{equation}
    P_{dyn}  = V\times I_{dyn}
\end{equation}
\begin{table}[htbp!]
\centering
\fontsize{8}{8}\selectfont{
\caption{Current consumption by Barnes workload with eight processes on real hardware Arm Cortex A-72.}
\label{tab:power-consumption}
\begin{tabular}{cccc}
\toprule
No. & Initial Current (A) & Final Current (A) & Difference (A) \\
\midrule
1  & 0.54 & 0.76 & 0.22 \\
2  & 0.61 & 1.07 & 0.46 \\
3  & 0.66 & 1.08 & 0.42 \\
4  & 0.65 & 0.90 & 0.25 \\
5  & 0.66 & 1.00 & 0.34 \\
6  & 0.60 & 1.00 & 0.40 \\
7  & 0.65 & 1.08 & 0.43 \\
8  & 0.66 & 1.07 & 0.41 \\
9  & 0.61 & 1.10 & 0.49 \\
10 & 0.66 & 0.98 & 0.32 \\
\bottomrule
\end{tabular}}
\end{table}
Multiple experiments were necessary to account for fluctuations in current due to external factors such as power cables and screen monitors, which continuously draw current. Therefore, averaging over several trials helped mitigate their impact and provided a more accurate measure of power consumption. 
The average power consumption on the physical Raspberry Pi 4 for a similar workload executed on gem5 using THEAS (\textbf{1.71613 W}) is calculated as \textbf{1.62 W}.
The experimental results demonstrate the effectiveness of the algorithm, revealing an average error of approximately \textbf{5.6\%}, which is the difference between the hardware metrics of Cortex A72 (1.62 W) and the simulation results of Gem5 (1.71613 W). An improvement of almost \textbf{4.8\%}   is gained over the standard gem5 setup from \textbf{10.4\%} to \textbf{5.6\%} in power consumption.  
\section{Conclusion}
In this research work, Task Heterogeneity Energy-Aware Scheduling (THEAS) is integrated and validated in Gem5, alongside an empirical power model and core-based resource allocation, in an ARM-based multicore ISA environment.  It addresses non-pinning workloads by assigning workloads to cores based on their computational demands, utilizing PMC event thresholds to manage the assignment.  The validation was carried out by comparing all the statistics produced through the Gem5 simulations with those calculated from the real hardware (ARM Cortex-A72). Considering the homogeneous architecture of the ARM with fewer workloads and frequency values, this work has the best results, with a difference of approximately \textbf{5.6\%} compared to the results based on real hardware. Further, almost \textbf{4.8\%} is improved
over the standard gem5 setup without performance degradation. Additionally, this is the first work to address the non-pinning environment of task execution in a multi-core environment with a small number of workload executions on a single chip. \par
While using non-pinning workloads, this research opens up new opportunities for future work, such as per-core DVFS and the integration of uncore (L3). In future work, we aim to explore the combined effects of THEAS and the empirical power model, addressing the opportunities mentioned earlier. The proposed approach will validate the applicability of THEAS in various real-world scenarios and offer more comprehensive insights into power consumption patterns across different workloads.

\vfill

\begin{thebibliography}{1}
\bibliographystyle{IEEEtran}

\bibitem{Jianwang}
Zhai, Jianwang, et al. `` McPAT-Calib: A microarchitecture power modeling framework for modern CPUs." \textit{IEEE/ACM International Conference On Computer Aided Design (ICCAD), 2021,} pp. 1-9.

\bibitem{Thasneema}
Thasneema, V., and Sreebha Bhaskaran. `` A Classification Model for Predicting Suitable Cache Level in a Multi-core Architecture." \textit{ International Conference on Communication, Control and Information Sciences (ICCISc), 2021, } pp. 1-6.

\bibitem{Brooks}
B. David, V. Tiwari, and M. Martonosi, `` Wattch: A framework for architectural-level power analysis and optimizations." \textit{ACM SIGARCH Computer Architecture News 28.2, 2000,} pp. 83-94.

\bibitem{Hebbar}
Hebbar, Ranjan, and Aleksandar Milenković. `` PMU-events-driven DVFS techniques for improving energy efficiency of modern processors." \textit{ACM Transactions on Modeling and Performance Evaluation of Computing Systems, 2022, } pp. 1-31.

\bibitem{Mazzola}
Mazzola, Sergio, et al. `` Data-Driven Power Modeling and Monitoring via Hardware Performance Counters Tracking." \textit{arXiv preprint arXiv:2401.01826,} 2024.

\bibitem{Zhai}
R. Bertran, M. Gonzalez, X. Martorell, N. Navarro, and E. Ayguade, `` McPAT-Calib: A RISC-V BOOM microarchitecture power modeling framework." \textit{IEEE Transactions on Computer-Aided Design of Integrated Circuits and Systems. 2022,} pp. 243-56.


\bibitem{MJ}
M. J. Walker. et al., `` Thermally-aware composite run-time CPU power models."  \textit{In 2016 26th International Workshop on Power and Timing Modeling, Optimization and Simulation (PATMOS) 2016},  pp. 17-24. 


\bibitem{Sergio}
S. Mazzola, T. Benz, B. Forsberg, and L. Benini,  `` A data-driven approach to lightweight dvfs-aware counter-based power modeling for heterogeneous platforms."  \textit{International Conference on Embedded Computer Systems. Cham: Springer International Publishing, 2022,} pp. pp. 346-361. 

\bibitem{Mahmoudi}
N. Mahmoudi, and H. Khazaei,  `` Performance modeling of serverless computing platforms." \textit{IEEE Transactions on Cloud Computing, 2020}, pp. 2834-2847.



\bibitem{Austin}
T. Austin, E. Larson, and D. Ernst, `` SimpleScalar: An infrastructure for computer system modeling." \textit{Computer. 2002}, pp. 59-67.

\bibitem{Patel}
A. Patel, F. Afram, and K. Ghose, `` Marss-x86: A qemu-based micro-architectural and systems simulator for x86 multicore processors." \textit{In 1st International Qemu Users’ Forum. IEEE, 2011,} pp. 29-30.

\bibitem{Binkert}
N. Binkert., et al.,  `` The Gem5 simulator." \textit{ ACM SIGARCH computer architecture news. 2011}, pp. 1-7.

\bibitem{Yassin}
Y. H. Yassin, M. Jahre, P. G. Kjeldsberg, S. Aunet, and F. Catthoor, `` Fast and accurate edge computing energy modeling and DVFS implementation in Gem5 using system call emulation mode." \textit{Journal of Signal Processing Systems 2021,} pp. 33-48.

\bibitem{Butko}
A. Butko. et al., `` Full-system simulation of big. little multicore architecture for performance and energy exploration." \textit{In 
 2016 IEEE 10th International Symposium on embedded multicore/many-core systems-on-chip (MCSOC). 2016}, pp. 201-208.

\bibitem{Reddy}
B. K.  Reddy. et al.,  `` Empirical CPU power modeling and estimation in the Gem5 simulator." \textit{In 2017 27th International Symposium on Power and Timing Modeling, Optimization and Simulation (PATMOS). 2017}, pp. 1-8.


\bibitem{Qiu}
Y. Qiu. et al.,   `` Performance Error Evaluation of Gem5 Simulator for ARM Server." \textit{In2023 IEEE 15th International Conference on ASIC (ASICON). IEEE,  2023, "} pp. 1-4.




\bibitem{Bischoff}
M. J.  Walker, S. Bischoff, S. Diestelhorst, G. Merrett, and B. Al-Hashimi, `` Hardware-validated CPU performance and energy modeling." \textit{In 2018 IEEE International Symposium on Performance Analysis of Systems and Software (ISPASS). 2018}, pp. 44-53.

 


\bibitem{NL}
Binkert, Dreslinski, et al., `` The M5 simulator: Modeling networked systems."  \textit{IEEE micro. 2006}, pp. 52-60.



\bibitem{Rabaey}
J. M. Rabaey. et al., `` Low power design methodologies." \textit{ Vol. 336. Springer Science \& Business Media, 2012}.



\bibitem{Arm_powermodel}
\url{https://www.Gem5.org/documentation/learning\_Gem5/part2/arm\_power\_modelling/.}


	
\end{thebibliography}
\end{document}